\documentclass[aps,prb,
twocolumn,showpacs]{revtex4}
\usepackage{t1enc}
\usepackage[latin1]{inputenc}
\usepackage[english]{babel}
\usepackage{graphics}
\usepackage{graphicx}
\usepackage{dcolumn}
\usepackage{bm}
\usepackage{color}
\usepackage{float}
\begin{document}
\title{Phase diagram of charge order in La$_{1.8-x}$Eu$_{0.2}$Sr$_x$CuO$_4$ from resonant soft x-ray diffraction}
\author{J\"org\ Fink,$^{1,2}$  Victor\ Soltwisch,$^1$ Jochen\ Geck,$^{2}$ Enrico\ Schierle,$^1$ Eugen\ Weschke,$^1$    Bernd\ B\"uchner,$^2$}
\affiliation{
$^1$Helmholtz-Zentrum Berlin für Materialien und Energie, Albert-Einstein-Str. 15, D-12489 Berlin, Germany\\
$^2$ Leibniz-Institute for Solid State and Materials Research Dresden, P.O.Box 270116, D-01171 Dresden, Germany\\}

\date{\today}

\begin{abstract}
Resonant soft x-ray scattering experiments with photon energies near the O $K$ and the Cu $L_3$ edge on the system La$_{1.8-x}$Eu$_{0.2}$Sr$_x$CuO$_4$ for $0.1\le x \le0.15$ are presented. A phase diagram for stripe-like charge ordering is obtained together with information on the structural transition into the low-temperature tetragonal phase. A clear dome for the charge ordering around $x=\frac 1 8$ is detected well below the structural transition. This result is quite different from other systems in which static stripes are detected. There the charge order is determined by the structural transition appearing at the same temperature. Furthermore we present results for the coherence length and the incommensurability of the stripe order as a function of Sr concentration.   
\end{abstract}
\pacs{ 61.05.cp, 71.45.Lr, 74.72.Dn, 75.50.Ee }
\maketitle
Soon after the discovery of high-T$_c$ superconductivity in doped cuprates, stripe-like phases, in which charge carrier poor antiferromagnetic domains are separated by charge carrier rich domain walls, have been discussed on the basis of a  Hartree-Fock analysis of the single-band Hubbard model.~\cite{Zaanen1989} Experimentally, static stripe order in cuprates has been first detected in La$_{1.6-x}$Nd$_{0.4}$Sr$_x$CuO$_4$ (LNSCO) near a doping concentration $x=\frac 1 8$ by elastic neutron scattering.~\cite{Tranquada1995} Later on static stripe order was also detected in the compounds La$_{2-x}$Ba$_x$CuO$_4$ (LBCO)~\cite{Fujita2002,Fujita2004,Abbamonte2005} and La$_{1.8-x}$Eu$_{0.2}$Sr$_x$CuO$_4$ (LESCO).~\cite{Klauss2000,Hucker2007,Fink2009} In all these systems it is believed that static stripe order is stabilized by a structural transition from a low temperature orthorhombic (LTO) to a low temperature tetragonal (LTT) phase in which the CuO$_6$ octahedra are tilted along the [110]$_{HTT}$ and [100]$_{HTT}$ directions of the high temperature tetragonal (HTT) phase, respectively. Generally, the stripe order is accompanied by a suppression of coherent superconductivity and the amount of this suppression is increasing with increasing tilt angle $\Phi$ in the LTT phase.~\cite{Buchner1994}
The tilt angle increases with decreasing ionic radius of substitutes on the La sites due to a chemical pressure along the CuO$_2$ layers. In LESCO with the small Eu ions the antiferromagnetic stripe order almost completely replaces the superconducting phase for $x\lesssim0.2$.
This apparent anticorrelation between stripe order and superconductivity, however, was recently questioned by the interpretation of transport data in LBCO in terms of a layer decoupled stripe superconductor with no phase coherence perpendicular to the CuO$_2$ layers.~\cite{Li2007}
So far complete phase diagrams for the structural, the charge, and the spin order in the systems  LBCO, LNSCO, and LESCO were proposed in Ref.~\onlinecite{Huecker2010}, Refs.~\onlinecite{Singer1999,Ichikawa2000}, and Ref.~\onlinecite{Klauss2000}, respectively. Possibly stripe like  order is not only restricted to La$_2$CuO$_4$ systems. A conceptually related
state may be the nematic one detected in underdoped YBa$_2$Cu$_3$O$_{7-\delta}$ (see Ref.~\onlinecite{Haug2010}), which only breaks the lattice rotation symmetry and
which may occur as an intermediate state upon melting of stripe order. The physical properties of stripe phases in doped cuprates have been recently reviewed in Refs.~\onlinecite{Kivelson2003,Vojta2009}.

In this Brief Report we complete our previous resonant soft x-ray scattering (RSXS) studies on the stripe like charge order in LESCO.~\cite{Fink2009} While there we have reported data only for doping concentrations $x\geq\frac 1 8$, in the present contribution we present measurements on $\it{both}$ sides of the concentration $x=\frac 1 8$. Furthermore we present more information on the
correlation length and on the incommensurability wave vector as a function of doping concentration. Finally we show an extended phase diagram for the lattice and the charge order in LESCO.

Traditionally, charge order was detected by measuring superstructure reflections with x-ray or neutron scattering. Both methods were used to study stripe-like charge order in LNSCO and LBCO.~\cite{Tranquada1995,Von1998,Huecker2010} In the case of LESCO, only hard x-ray scattering was successful.~\cite{Cyr-Choiniere2009} RSXS at the O $K$ and Cu $L$ edges is a particular sensitive method to detect charge order in doped cuprates.~\cite{Abbamonte2005,Fink2009} At the pre-peak of the O $K$ edge, the form factor for a hole is enhanced by a factor of 82,  which provides a particular sensitivity to modulations of the hole density.~\cite{Abbamonte2005} This is important in view of the fact that conventional x-ray scattering probes lattice distortions that could alternatively also be induced by normal structural transitions or spin-Peierls transitions.~\cite{Hase1993} A considerable resonant enhancement of the charge-order diffraction peak is also observed at the Cu $L$ edge, but according to Ref.~\onlinecite{Abbamonte2006}, predominantly lattice distortions are probed there. Finally, we point out that RSXS in the present case is not a surface sensitive technique. Our estimate for the probing depths $d$ in LESCO yields at the pre-peak of the O $K$ edge and at the white line of the Cu $L_3$ edge, $d$= 75 nm and $d$= 45 nm, respectively.

The RSXS experiments were performed at the BESSY undulator beamline UE 46-PGM using a two circle UHV diffractometer. More experimental details were presented in our previous paper.~\cite{Fink2009} We denote the wave vector $\textbf{Q}  =
(2\pi h/a\ 2\pi k/b\ 2\pi l/c)$ with Miller indices $(h k l)$
where in the LTT phase $a = b$ = 3.79 {\AA}  and $c$ = 13.14 {\AA} for $x$ = 0.1 - 0.15.

\begin{figure}[t]
\includegraphics[width=7 cm]{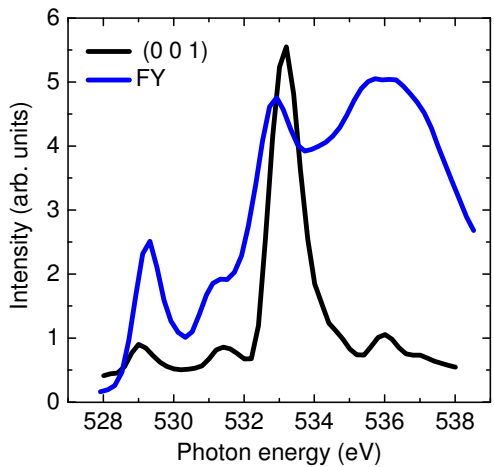}
\caption{(Color online) X-ray absorption spectrum of La$_{1.65}$Eu$_{0.2}$Sr$_{0.15}$CuO$_4$ near the O $K$ edge measured in the 
fluorescence yield mode (FY) in light gray (blue). Also shown in black is the (001) superstructure reflection intensity as a function of the photon energy.}
\label{xas}
\end{figure}

In Fig.~\ref{xas} we present x-ray absorption spectroscopy (XAS) 
measurements using the fluorescence method near the O $K$  edge for a LESCO ($x$ = 0.15)
single crystal at a temperature T = 6 K.  
In accordance with previous electron energy-loss spectroscopy and 
XAS studies\cite{Romberg1990,Chen1991} we see at 529.5 eV transitions into the O $2p$ doped hole states 
in the conduction band and at
531.5 eV transitions into the Cu $3d$ upper Hubbard band hybridized with O $2p$ states. Near 533 eV we realize a further peak due to a hybridization of O $2p$ states with rare earth (RE=La and Eu) $5d$ and/or $4f$ states.~\cite{Nucker1988} 
In Fig.~\ref{xas} we also show the photon energy dependence of the (001) reflection 
measured in the LTT phase at 6K. In this phase, neighboring CuO$_2$ planes 
are rotated by 90$^\circ$, yielding O sites with different (rotated) local 
environments. As a result, the (001) reflection becomes allowed at 
resonance.~\cite{Dmitrienko1983} The strong resonance at 533.5eV is due to 
octahedral tilts, which cause different local environments and affect 
the hybridization between the apical O and the RE orbitals. In the LTO phase, 
neighboring CuO$_2$ planes are just shifted, not rotated, with respect to 
each other. In this case the (001) reflection remains forbidden even at 
resonance. In this context we remark that the resonances detected for this reflection for the two pre-peak energies is much smaller. Using the resonance at 533.5 eV it is possible the detect the LTO-LTT phase transition with soft x-rays. This transition is not detectable with soft x-ray diffraction by the orthorhombic strain (a-b splitting) because the (100)/(010) reflections cannot be reached due to the limited range in \textbf{Q} space.

\begin{figure}[t]
\includegraphics[width=8 cm]{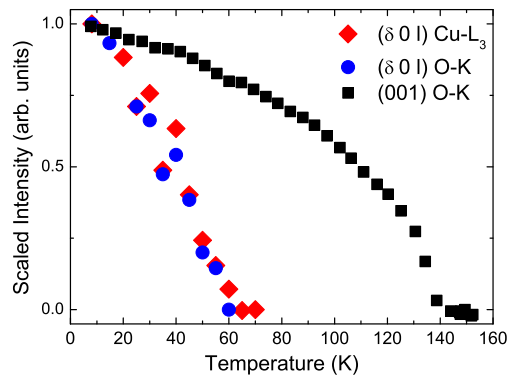}
\caption{(Color online) Temperature dependence 
of the intensities of the superstructure
reflections of La$_{1.65}$Eu$_{0.2}$Sr$_{0.15}$CuO$_4$ normalized to the intensity at $T$=6 K. Squares: (001) reflection measured with photon energies h$\nu$= 533.2 eV near the O $K.
$ edge. Circles: (0.254 0 0.75) reflection measured with photon energies h$\nu$= 529.2 eV near the O $K$ edge. Diamonds: (0.254 0 1.6) reflection measured with photon energies h$\nu$= 929.8 eV near the Cu $L_3$ edge }
\label{tdep}
\end{figure}

\begin{figure}
\includegraphics[width=8 cm]{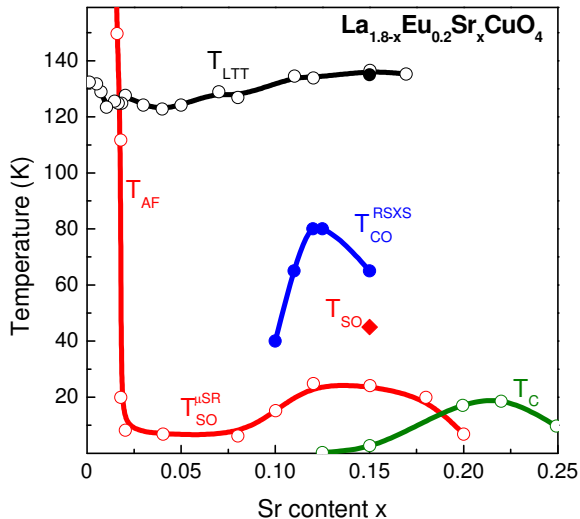}
\caption{(Color online) Phase diagram of La$_{1.8-x}$Eu$_{0.2}$Sr$_x$CuO$_4$ showing transition temperatures for the LTT phase $T_{LTT}$, the antiferromagnetic structure $T_{AF}$, the magnetic stripe order $T_{SO}$, the stripe like charge order $T_{CO}$, and the superconducting transition temperature $T_{C}$. Closed circles from the present RSXS experiments. Open circles from Ref.~\onlinecite{Klauss2000}. Closed diamond from neutron diffraction data presented in Ref.~\onlinecite{Hucker2007}.}
\label{phase}
\end{figure}

In Fig.~\ref{tdep} we present the temperature dependence of the intensity of the (001) reflection for a LESCO ($x$=0.15) crystal.
A sudden rise near T$_{LTT}$=135 K indicates a nearly first order like transition into the LTT phase. This T$_{LTT}$ value nicely agrees with the x-ray diffraction value~\cite{Klauss2000} and the $^{63}$Cu nuclear magnetic resonance (NQR) value~\cite{Simovic2003} (see also Fig.~\ref{phase}). The small dip near $T$= 60 K possibly signals a weak coupling between the structural order parameter and the charge order parameter. We compare in Fig.~\ref{tdep} these data with the temperature dependence of the superstructure reflection related to stripe like charge order measured at the O $K$ edge and the Cu $L_3$ edge which are similar to those presented in our previous publication.~\cite{Fink2009} The rapid decay of the intensity with increasing temperature raises the question, whether this charge order is dominated by disorder or by slow fluctuations. We emphasize that the present data were all taken from the same samples using the same diffractometer.         
Within error bars we see no difference in the temperature dependence of the superstructure intensities measured at the O $K$ edge and the Cu $L_3$ edge. Remembering that the probing depths at the two edges is different by a factor of two, this signals that the charge order detected in this material is not a surface effect. Extrapolating the intensities to zero yields a charge order temperature $T_{CO}$=65 K in agreement with our previous results. Similar measurements were performed also for smaller Sr concentrations. The charge order temperatures $T_{CO}$ are presented in a phase diagram for LESCO
shown in Fig.~\ref{phase}. We note that for x=0.125 our $T_{CO}$ value is in perfect agreement with the hard x-ray value from Ref.~\onlinecite{Cyr-Choiniere2009}, again indicating that surface effects are not important in our RSXS study. In Fig.~\ref{phase}, we have included data on $T_{LTT}$, the transition temperature $T_{AF}$ into the antiferromagnetic order at low Sr concentrations, the transition temperature  
$T_{SO}^{\mu SR}$ for the magnetic stripe order derived from $\mu$SR, and the superconducting transition temperature $T_c$, all taken from Ref.~\onlinecite{Klauss2000}. Furthermore, we have included the $T_{LTT}$ value for x=0.15 derived from the present RSXS investigation. Finally we have also added a value for the magnetic stripe order transition temperature $T_{SO}$ derived from neutron diffraction for x=0.15.~\cite{Hucker2007}
The phase boundary of charge order in LESCO exhibits a clear dome-like shape around  $x=\frac 1 8$. It can be observed here in detail without interference by the structural LTT phase transition that largely determines the behavior in the cases of LBCO and LNSCO. A similar shape is also found for the magnetic phase transition probed by  $\mu$SR. The data point for magnetic stripe-like order from neutron diffraction for $x$=0.15 indicates that there is a large temperature range where charge order exists without spin order, as also found for LBCO and LNSCO.  From many previous studies on static stripe phases it is clear that the stripes are stabilized by the corrugation of the CuO$_2$ layers in the LTT phase. LESCO is the first example in which the lattice transition temperature $T_{LTT}$ is so high that the charge order cannot be stabilized any more at this temperature and therefore a gap of 55 K exists between $T_{LTT}$ and $T_{CO}$. This is a remarkable result since it demonstrates that the charge order is purely driven by entropy.

\begin{figure}[t]
\includegraphics[width=7 cm]{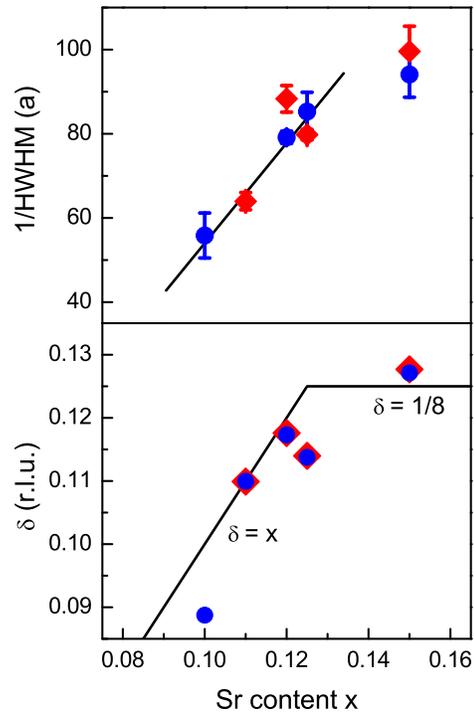}
\caption{(Color online) Upper panel: concentration dependence of the coherence length in units of the lattice constant $a$ of static stripe order in La$_{1.8-x}$Eu$_{0.2}$Sr$_x$CuO$_4$. Lower panel: concentration dependence of the incommensurability wave vector $\delta$ for the stripe order which is one half of the wave vector for charge order. The assignment to the symbols is the same as in Fig. 2}
\label{delta}
\end{figure}

The widths of the diffraction profiles measured at low temperatures is about 5 times larger than the instrumental resolution. This points toward disorder effects and/or glassy behavior. From the widths we have determined the coherence lengths of the charge order. These lengths are of the order of about 100 lattice constants, $a$, and increase, at least for smaller $x$, linearly with increasing Sr concentration (see Fig.~\ref{delta}). The increase of the coherence length with increasing doping level in this range indicates that the length scale of the charge order is not determined by the impurity potential of the doping atoms. Instead it clearly indicates that the coherence length is dominated by an electronic mechanism. The behavior suggests a connection to the optimum doping level of $x=\frac 1 8$. This would imply a decrease of the coherence length at higher Sr concentrations, which, however, is not observed, albeit there is a clear deviation from the linear increase for $x$=0.15.  
It is also interesting to note that our recent RSXS data on LBCO~\cite{Wilkins2010} indicate that in this system the coherence lengths are about a factor 3 larger than in LESCO. This result can be explained by the fact that in LESCO additional disorder is induced by replacing 20 \% of the La atoms by Eu. This points towards 
the fact that also structural parameters play a role for the charge order. The result also signals that it is much easier to detect static stripe order using RSXS on LBCO than on LESCO since the intensity of the superstructure reflection is determined by the squared coherence length.

In Fig.~\ref{delta} we also show the incommensurability vector $\delta$ for the stripe order, which is twice that vector for the charge order, as a function of $x$.  For $x\leq\frac 1 8$
the wave vector $\delta$ is determined by the Sr concentration, i.e. $\delta=x$. For concentrations larger than $\frac 1 8$, the wave vectors saturates at $\frac 1 8$. A similar behavior is also found for the other systems (LBCO and LNSCO) showing static stripe like order and also for dynamic correlations in La$_{1-x}$Sr$_x$CuO$_4$.~\cite{Yamada1998} As pointed out previously, this result clearly signals, that the charge order in these systems is not a conventional charge density wave caused by nesting since the distance between the antinodal Fermi surface decreases with increasing Sr concentration. Thus these results point to more strong coupling scenarios for the stripe formation.

In conclusion we have presented a complete phase diagram of charge order in La$_{1.8-x}$Eu$_{0.2}$Sr$_x$CuO$_4$ and we have compiled this charge order with other transition temperatures of magnetic (stripe) order, structural transition temperatures, and values for superconducting $T_c$. It turned out that LESCO is the only system among the compounds showing static stripe like order, in which the charge order is not directly determined by the existence of a low temperature tetragonal structure. This means LESCO displays well separated phase transitions with $T_{SO}<T_{CO}<T_{LTT}$. A further result is that the coherence length for the charge modulation is not determined by the impurity potential of the doping atoms. It seems more plausible to relate the coherence lengths to the proximity to the commensurate stripe like structure near for $x=\frac 1 8$. Our doping dependent study provides clear evidence that the charge 
ordering in LESCO is largely determined by electronic mechanisms. But 
the comparison to LBCO also indicates that structural parameters play some role.
Finally we have found a new resonance in RSXS experiments near the O $K$ edge which allows to detect the structural transition between the LTO and the LTT phase.

Financial support by the DFG through FOR 538 is appreciated by the IFW based researchers. J.G. acknowledges the support by the DFG through the Emmy-Noether program. We thank C. Sch\"u{\ss}ler-Langeheine for valuable discussions and making available his equipment.

\bibliography{CO_phasedia}

\begin{thebibliography}{27}
\expandafter\ifx\csname natexlab\endcsname\relax\def\natexlab#1{#1}\fi
\expandafter\ifx\csname bibnamefont\endcsname\relax
  \def\bibnamefont#1{#1}\fi
\expandafter\ifx\csname bibfnamefont\endcsname\relax
  \def\bibfnamefont#1{#1}\fi
\expandafter\ifx\csname citenamefont\endcsname\relax
  \def\citenamefont#1{#1}\fi
\expandafter\ifx\csname url\endcsname\relax
  \def\url#1{\texttt{#1}}\fi
\expandafter\ifx\csname urlprefix\endcsname\relax\def\urlprefix{URL }\fi
\providecommand{\bibinfo}[2]{#2}
\providecommand{\eprint}[2][]{\url{#2}}

\bibitem[{\citenamefont{Zaanen and Gunnarsson}(1989)}]{Zaanen1989}
\bibinfo{author}{\bibfnamefont{J.}~\bibnamefont{Zaanen}} \bibnamefont{and}
  \bibinfo{author}{\bibfnamefont{O.}~\bibnamefont{Gunnarsson}},
  \bibinfo{journal}{Phys. Rev. B} \textbf{\bibinfo{volume}{40}},
  \bibinfo{pages}{7391} (\bibinfo{year}{1989}).

\bibitem[{\citenamefont{Tranquada et~al.}(1995)\citenamefont{Tranquada,
  Sternlieb, Axe, Nakamura, and Uchida}}]{Tranquada1995}
\bibinfo{author}{\bibfnamefont{J.~M.} \bibnamefont{Tranquada}},
  \bibinfo{author}{\bibfnamefont{B.~J.} \bibnamefont{Sternlieb}},
  \bibinfo{author}{\bibfnamefont{J.~D.} \bibnamefont{Axe}},
  \bibinfo{author}{\bibfnamefont{Y.}~\bibnamefont{Nakamura}}, \bibnamefont{and}
  \bibinfo{author}{\bibfnamefont{S.}~\bibnamefont{Uchida}},
  \bibinfo{journal}{Nature} \textbf{\bibinfo{volume}{375}},
  \bibinfo{pages}{561} (\bibinfo{year}{1995}).

\bibitem[{\citenamefont{Fujita et~al.}(2002)\citenamefont{Fujita, Goka, Yamada,
  and Matsuda}}]{Fujita2002}
\bibinfo{author}{\bibfnamefont{M.}~\bibnamefont{Fujita}},
  \bibinfo{author}{\bibfnamefont{H.}~\bibnamefont{Goka}},
  \bibinfo{author}{\bibfnamefont{K.}~\bibnamefont{Yamada}}, \bibnamefont{and}
  \bibinfo{author}{\bibfnamefont{M.}~\bibnamefont{Matsuda}},
  \bibinfo{journal}{Phys. Rev. Lett.} \textbf{\bibinfo{volume}{88}},
  \bibinfo{pages}{167008} (\bibinfo{year}{2002}).

\bibitem[{\citenamefont{Fujita et~al.}(2004)\citenamefont{Fujita, Goka, Yamada,
  Tranquada, and Regnault}}]{Fujita2004}
\bibinfo{author}{\bibfnamefont{M.}~\bibnamefont{Fujita}},
  \bibinfo{author}{\bibfnamefont{H.}~\bibnamefont{Goka}},
  \bibinfo{author}{\bibfnamefont{K.}~\bibnamefont{Yamada}},
  \bibinfo{author}{\bibfnamefont{J.~M.} \bibnamefont{Tranquada}},
  \bibnamefont{and} \bibinfo{author}{\bibfnamefont{L.~P.}
  \bibnamefont{Regnault}}, \bibinfo{journal}{Phys. Rev. B}
  \textbf{\bibinfo{volume}{70}}, \bibinfo{pages}{104517}
  (\bibinfo{year}{2004}).

\bibitem[{\citenamefont{Abbamonte et~al.}(2005)\citenamefont{Abbamonte, Rusydi,
  Smadici, Gu, Sawatzky, and Feng}}]{Abbamonte2005}
\bibinfo{author}{\bibfnamefont{P.}~\bibnamefont{Abbamonte}},
  \bibinfo{author}{\bibfnamefont{A.}~\bibnamefont{Rusydi}},
  \bibinfo{author}{\bibfnamefont{S.}~\bibnamefont{Smadici}},
  \bibinfo{author}{\bibfnamefont{G.~D.} \bibnamefont{Gu}},
  \bibinfo{author}{\bibfnamefont{G.~A.} \bibnamefont{Sawatzky}},
  \bibnamefont{and} \bibinfo{author}{\bibfnamefont{D.~L.} \bibnamefont{Feng}},
  \bibinfo{journal}{Nature Physics} \textbf{\bibinfo{volume}{1}},
  \bibinfo{pages}{155} (\bibinfo{year}{2005}).

\bibitem[{\citenamefont{Klauss et~al.}(2000)\citenamefont{Klauss, Wagener,
  Hillberg, Kopmann, Walf, Litterst, Hucker, and Buchner}}]{Klauss2000}
\bibinfo{author}{\bibfnamefont{H.~H.} \bibnamefont{Klauss}},
  \bibinfo{author}{\bibfnamefont{W.}~\bibnamefont{Wagener}},
  \bibinfo{author}{\bibfnamefont{M.}~\bibnamefont{Hillberg}},
  \bibinfo{author}{\bibfnamefont{W.}~\bibnamefont{Kopmann}},
  \bibinfo{author}{\bibfnamefont{H.}~\bibnamefont{Walf}},
  \bibinfo{author}{\bibfnamefont{F.~J.} \bibnamefont{Litterst}},
  \bibinfo{author}{\bibfnamefont{M.}~\bibnamefont{Hucker}}, \bibnamefont{and}
  \bibinfo{author}{\bibfnamefont{B.}~\bibnamefont{Buchner}},
  \bibinfo{journal}{Phys. Rev. Lett.} \textbf{\bibinfo{volume}{85}},
  \bibinfo{pages}{4590} (\bibinfo{year}{2000}).

\bibitem[{\citenamefont{Hucker et~al.}(2007)\citenamefont{Hucker, Gu,
  Tranquada, von Zimmermann, Klauss, Curro, Braden, and Buchner}}]{Hucker2007}
\bibinfo{author}{\bibfnamefont{M.}~\bibnamefont{Hucker}},
  \bibinfo{author}{\bibfnamefont{G.~D.} \bibnamefont{Gu}},
  \bibinfo{author}{\bibfnamefont{J.~M.} \bibnamefont{Tranquada}},
  \bibinfo{author}{\bibfnamefont{M.}~\bibnamefont{von Zimmermann}},
  \bibinfo{author}{\bibfnamefont{H.~H.} \bibnamefont{Klauss}},
  \bibinfo{author}{\bibfnamefont{N.~J.} \bibnamefont{Curro}},
  \bibinfo{author}{\bibfnamefont{M.}~\bibnamefont{Braden}}, \bibnamefont{and}
  \bibinfo{author}{\bibfnamefont{B.}~\bibnamefont{Buchner}},
  \bibinfo{journal}{Physica C (Amsterdam, Neth.)}
  \textbf{\bibinfo{volume}{460}}, \bibinfo{pages}{170} (\bibinfo{year}{2007}).

\bibitem[{\citenamefont{Fink et~al.}(2009)\citenamefont{Fink, Schierle,
  Weschke, Geck, Hawthorn, Soltwisch, Wadati, Wu, D\"urr, Wizent
  et~al.}}]{Fink2009}
\bibinfo{author}{\bibfnamefont{J.}~\bibnamefont{Fink}},
  \bibinfo{author}{\bibfnamefont{E.}~\bibnamefont{Schierle}},
  \bibinfo{author}{\bibfnamefont{E.}~\bibnamefont{Weschke}},
  \bibinfo{author}{\bibfnamefont{J.}~\bibnamefont{Geck}},
  \bibinfo{author}{\bibfnamefont{D.}~\bibnamefont{Hawthorn}},
  \bibinfo{author}{\bibfnamefont{V.}~\bibnamefont{Soltwisch}},
  \bibinfo{author}{\bibfnamefont{H.}~\bibnamefont{Wadati}},
  \bibinfo{author}{\bibfnamefont{H.-H.} \bibnamefont{Wu}},
  \bibinfo{author}{\bibfnamefont{H.~A.} \bibnamefont{D\"urr}},
  \bibinfo{author}{\bibfnamefont{N.}~\bibnamefont{Wizent}},
  \bibnamefont{et~al.}, \bibinfo{journal}{Phys. Rev. B}
  \textbf{\bibinfo{volume}{79}}, \bibinfo{pages}{100502}
  (\bibinfo{year}{2009}).

\bibitem[{\citenamefont{B\"uchner et~al.}(1994)\citenamefont{B\"uchner, Breuer,
  Freimuth, and Kampf}}]{Buchner1994}
\bibinfo{author}{\bibfnamefont{B.}~\bibnamefont{B\"uchner}},
  \bibinfo{author}{\bibfnamefont{M.}~\bibnamefont{Breuer}},
  \bibinfo{author}{\bibfnamefont{A.}~\bibnamefont{Freimuth}}, \bibnamefont{and}
  \bibinfo{author}{\bibfnamefont{A.~P.} \bibnamefont{Kampf}},
  \bibinfo{journal}{Phys. Rev. Lett.} \textbf{\bibinfo{volume}{73}},
  \bibinfo{pages}{1841} (\bibinfo{year}{1994}).

\bibitem[{\citenamefont{Li et~al.}(2007)\citenamefont{Li, Hucker, Gu, Tsvelik,
  and Tranquada}}]{Li2007}
\bibinfo{author}{\bibfnamefont{Q.}~\bibnamefont{Li}},
  \bibinfo{author}{\bibfnamefont{M.}~\bibnamefont{Hucker}},
  \bibinfo{author}{\bibfnamefont{G.~D.} \bibnamefont{Gu}},
  \bibinfo{author}{\bibfnamefont{A.~M.} \bibnamefont{Tsvelik}},
  \bibnamefont{and} \bibinfo{author}{\bibfnamefont{J.~M.}
  \bibnamefont{Tranquada}}, \bibinfo{journal}{Phys. Rev. Lett.}
  \textbf{\bibinfo{volume}{99}}, \bibinfo{pages}{067001}
  (\bibinfo{year}{2007}).

\bibitem[{\citenamefont{Huecker et~al.}(2010)\citenamefont{Huecker,
  v~Zimmermann, Gu, Xu, Wen, Xu, Kang, Zheludev, and Tranquada}}]{Huecker2010}
\bibinfo{author}{\bibfnamefont{M.}~\bibnamefont{Huecker}},
  \bibinfo{author}{\bibfnamefont{M.}~\bibnamefont{v~Zimmermann}},
  \bibinfo{author}{\bibfnamefont{G.~D.} \bibnamefont{Gu}},
  \bibinfo{author}{\bibfnamefont{Z.~J.} \bibnamefont{Xu}},
  \bibinfo{author}{\bibfnamefont{J.~S.} \bibnamefont{Wen}},
  \bibinfo{author}{\bibfnamefont{G.}~\bibnamefont{Xu}},
  \bibinfo{author}{\bibfnamefont{H.~J.} \bibnamefont{Kang}},
  \bibinfo{author}{\bibfnamefont{A.}~\bibnamefont{Zheludev}}, \bibnamefont{and}
  \bibinfo{author}{\bibfnamefont{J.~M.} \bibnamefont{Tranquada}},
  \bibinfo{journal}{arXiv:1005.5191}  (\bibinfo{year}{2010}).

\bibitem[{\citenamefont{Singer et~al.}(1999)\citenamefont{Singer, Hunt,
  Cederstr\"om, and Imai}}]{Singer1999}
\bibinfo{author}{\bibfnamefont{P.~M.} \bibnamefont{Singer}},
  \bibinfo{author}{\bibfnamefont{A.~W.} \bibnamefont{Hunt}},
  \bibinfo{author}{\bibfnamefont{A.~F.} \bibnamefont{Cederstr\"om}},
  \bibnamefont{and} \bibinfo{author}{\bibfnamefont{T.}~\bibnamefont{Imai}},
  \bibinfo{journal}{Phys. Rev. B} \textbf{\bibinfo{volume}{60}},
  \bibinfo{pages}{15345} (\bibinfo{year}{1999}).

\bibitem[{\citenamefont{Ichikawa et~al.}(2000)\citenamefont{Ichikawa, Uchida,
  Tranquada, Niemoller, Gehring, Lee, and Schneider}}]{Ichikawa2000}
\bibinfo{author}{\bibfnamefont{N.}~\bibnamefont{Ichikawa}},
  \bibinfo{author}{\bibfnamefont{S.}~\bibnamefont{Uchida}},
  \bibinfo{author}{\bibfnamefont{J.}~\bibnamefont{Tranquada}},
  \bibinfo{author}{\bibfnamefont{T.}~\bibnamefont{Niemoller}},
  \bibinfo{author}{\bibfnamefont{P.}~\bibnamefont{Gehring}},
  \bibinfo{author}{\bibfnamefont{S.}~\bibnamefont{Lee}}, \bibnamefont{and}
  \bibinfo{author}{\bibfnamefont{J.}~\bibnamefont{Schneider}},
  \bibinfo{journal}{Phys. Rev. Lett.} \textbf{\bibinfo{volume}{85}},
  \bibinfo{pages}{1738} (\bibinfo{year}{2000}).

\bibitem[{\citenamefont{Haug et~al.}(2010)\citenamefont{Haug, Hinkov, Sidis,
  Bourges, Christensen, Ivanov, Keller, Lin, and Keimer}}]{Haug2010}
\bibinfo{author}{\bibfnamefont{D.}~\bibnamefont{Haug}},
  \bibinfo{author}{\bibfnamefont{V.}~\bibnamefont{Hinkov}},
  \bibinfo{author}{\bibfnamefont{Y.}~\bibnamefont{Sidis}},
  \bibinfo{author}{\bibfnamefont{P.}~\bibnamefont{Bourges}},
  \bibinfo{author}{\bibfnamefont{N.~B.} \bibnamefont{Christensen}},
  \bibinfo{author}{\bibfnamefont{A.}~\bibnamefont{Ivanov}},
  \bibinfo{author}{\bibfnamefont{T.}~\bibnamefont{Keller}},
  \bibinfo{author}{\bibfnamefont{C.~T.} \bibnamefont{Lin}}, \bibnamefont{and}
  \bibinfo{author}{\bibfnamefont{B.}~\bibnamefont{Keimer}},
  \bibinfo{journal}{arXiv:1008.4298}  (\bibinfo{year}{2010}).

\bibitem[{\citenamefont{Kivelson et~al.}(2003)\citenamefont{Kivelson, Bindloss,
  Fradkin, Oganesyan, Tranquada, Kapitulnik, and Howald}}]{Kivelson2003}
\bibinfo{author}{\bibfnamefont{S.~A.} \bibnamefont{Kivelson}},
  \bibinfo{author}{\bibfnamefont{I.~P.} \bibnamefont{Bindloss}},
  \bibinfo{author}{\bibfnamefont{E.}~\bibnamefont{Fradkin}},
  \bibinfo{author}{\bibfnamefont{V.}~\bibnamefont{Oganesyan}},
  \bibinfo{author}{\bibfnamefont{J.~M.} \bibnamefont{Tranquada}},
  \bibinfo{author}{\bibfnamefont{A.}~\bibnamefont{Kapitulnik}},
  \bibnamefont{and} \bibinfo{author}{\bibfnamefont{C.}~\bibnamefont{Howald}},
  \bibinfo{journal}{Reviews of Modern Physics} \textbf{\bibinfo{volume}{75}},
  \bibinfo{pages}{1201} (\bibinfo{year}{2003}).

\bibitem[{\citenamefont{Vojta}(2009)}]{Vojta2009}
\bibinfo{author}{\bibfnamefont{M.}~\bibnamefont{Vojta}},
  \bibinfo{journal}{Advances in Physics} \textbf{\bibinfo{volume}{58}},
  \bibinfo{pages}{699} (\bibinfo{year}{2009}).

\bibitem[{\citenamefont{Von~Zimmermann
  et~al.}(1998)\citenamefont{Von~Zimmermann, Vigliante, Niemoller, Ichikawa,
  Frello, Madsen, Wochner, Uchida, Andersen, Tranquada et~al.}}]{Von1998}
\bibinfo{author}{\bibfnamefont{M.}~\bibnamefont{Von~Zimmermann}},
  \bibinfo{author}{\bibfnamefont{A.}~\bibnamefont{Vigliante}},
  \bibinfo{author}{\bibfnamefont{T.}~\bibnamefont{Niemoller}},
  \bibinfo{author}{\bibfnamefont{N.}~\bibnamefont{Ichikawa}},
  \bibinfo{author}{\bibfnamefont{T.}~\bibnamefont{Frello}},
  \bibinfo{author}{\bibfnamefont{J.}~\bibnamefont{Madsen}},
  \bibinfo{author}{\bibfnamefont{P.}~\bibnamefont{Wochner}},
  \bibinfo{author}{\bibfnamefont{S.}~\bibnamefont{Uchida}},
  \bibinfo{author}{\bibfnamefont{N.~H.} \bibnamefont{Andersen}},
  \bibinfo{author}{\bibfnamefont{J.~M.} \bibnamefont{Tranquada}},
  \bibnamefont{et~al.}, \bibinfo{journal}{Europhysics Letters}
  \textbf{\bibinfo{volume}{41}}, \bibinfo{pages}{629} (\bibinfo{year}{1998}).

\bibitem[{\citenamefont{Cyr-Choiniere et~al.}(2009)\citenamefont{Cyr-Choiniere,
  Daou, Laliberte, LeBoeuf, Doiron-Leyraud, Chang, Yan, Cheng, Zhou, Goodenough
  et~al.}}]{Cyr-Choiniere2009}
\bibinfo{author}{\bibfnamefont{O.}~\bibnamefont{Cyr-Choiniere}},
  \bibinfo{author}{\bibfnamefont{R.}~\bibnamefont{Daou}},
  \bibinfo{author}{\bibfnamefont{F.}~\bibnamefont{Laliberte}},
  \bibinfo{author}{\bibfnamefont{D.}~\bibnamefont{LeBoeuf}},
  \bibinfo{author}{\bibfnamefont{N.}~\bibnamefont{Doiron-Leyraud}},
  \bibinfo{author}{\bibfnamefont{J.}~\bibnamefont{Chang}},
  \bibinfo{author}{\bibfnamefont{J.-Q.} \bibnamefont{Yan}},
  \bibinfo{author}{\bibfnamefont{J.-G.} \bibnamefont{Cheng}},
  \bibinfo{author}{\bibfnamefont{J.-S.} \bibnamefont{Zhou}},
  \bibinfo{author}{\bibfnamefont{J.~B.} \bibnamefont{Goodenough}},
  \bibnamefont{et~al.}, \bibinfo{journal}{Nature}
  \textbf{\bibinfo{volume}{458}}, \bibinfo{pages}{743} (\bibinfo{year}{2009}).

\bibitem[{\citenamefont{Hase et~al.}(1993)\citenamefont{Hase, Terasaki, and
  Uchinokura}}]{Hase1993}
\bibinfo{author}{\bibfnamefont{M.}~\bibnamefont{Hase}},
  \bibinfo{author}{\bibfnamefont{I.}~\bibnamefont{Terasaki}}, \bibnamefont{and}
  \bibinfo{author}{\bibfnamefont{K.}~\bibnamefont{Uchinokura}},
  \bibinfo{journal}{Phys. Rev. Lett.} \textbf{\bibinfo{volume}{70}},
  \bibinfo{pages}{3651} (\bibinfo{year}{1993}).

\bibitem[{\citenamefont{Abbamonte}(2006)}]{Abbamonte2006}
\bibinfo{author}{\bibfnamefont{P.}~\bibnamefont{Abbamonte}},
  \bibinfo{journal}{Phys. Rev. B} \textbf{\bibinfo{volume}{74}},
  \bibinfo{pages}{195113} (\bibinfo{year}{2006}).

\bibitem[{\citenamefont{Romberg et~al.}(1990)\citenamefont{Romberg, Alexander,
  Nucker, Adelmann, and Fink}}]{Romberg1990}
\bibinfo{author}{\bibfnamefont{H.}~\bibnamefont{Romberg}},
  \bibinfo{author}{\bibfnamefont{M.}~\bibnamefont{Alexander}},
  \bibinfo{author}{\bibfnamefont{N.}~\bibnamefont{Nucker}},
  \bibinfo{author}{\bibfnamefont{P.}~\bibnamefont{Adelmann}}, \bibnamefont{and}
  \bibinfo{author}{\bibfnamefont{J.}~\bibnamefont{Fink}},
  \bibinfo{journal}{Phys. Rev. B} \textbf{\bibinfo{volume}{42}},
  \bibinfo{pages}{8768} (\bibinfo{year}{1990}).

\bibitem[{\citenamefont{Chen et~al.}(1991)\citenamefont{Chen, Sette, Ma,
  Hybertsen, Stechel, Foulkes, Schluter, Cheong, Cooper, Rupp
  et~al.}}]{Chen1991}
\bibinfo{author}{\bibfnamefont{C.~T.} \bibnamefont{Chen}},
  \bibinfo{author}{\bibfnamefont{F.}~\bibnamefont{Sette}},
  \bibinfo{author}{\bibfnamefont{Y.}~\bibnamefont{Ma}},
  \bibinfo{author}{\bibfnamefont{M.~S.} \bibnamefont{Hybertsen}},
  \bibinfo{author}{\bibfnamefont{E.~B.} \bibnamefont{Stechel}},
  \bibinfo{author}{\bibfnamefont{W.~M.~C.} \bibnamefont{Foulkes}},
  \bibinfo{author}{\bibfnamefont{M.}~\bibnamefont{Schluter}},
  \bibinfo{author}{\bibfnamefont{S.~W.} \bibnamefont{Cheong}},
  \bibinfo{author}{\bibfnamefont{A.~S.} \bibnamefont{Cooper}},
  \bibinfo{author}{\bibfnamefont{L.~W.} \bibnamefont{Rupp}},
  \bibnamefont{et~al.}, \bibinfo{journal}{Phys. Rev. Lett.}
  \textbf{\bibinfo{volume}{66}}, \bibinfo{pages}{104} (\bibinfo{year}{1991}).

\bibitem[{\citenamefont{N\"ucker et~al.}(1988)\citenamefont{N\"ucker, Fink,
  Fuggle, Durham, and Temmerman}}]{Nucker1988}
\bibinfo{author}{\bibfnamefont{N.}~\bibnamefont{N\"ucker}},
  \bibinfo{author}{\bibfnamefont{J.}~\bibnamefont{Fink}},
  \bibinfo{author}{\bibfnamefont{J.~C.} \bibnamefont{Fuggle}},
  \bibinfo{author}{\bibfnamefont{P.~J.} \bibnamefont{Durham}},
  \bibnamefont{and} \bibinfo{author}{\bibfnamefont{W.~M.}
  \bibnamefont{Temmerman}}, \bibinfo{journal}{Phys. Rev. B}
  \textbf{\bibinfo{volume}{37}}, \bibinfo{pages}{5158} (\bibinfo{year}{1988}).

\bibitem[{\citenamefont{Dmitrienko}(1983)}]{Dmitrienko1983}
\bibinfo{author}{\bibfnamefont{V.}~\bibnamefont{Dmitrienko}},
  \bibinfo{journal}{Acta Cryst.} \textbf{\bibinfo{volume}{A39}},
  \bibinfo{pages}{29} (\bibinfo{year}{1983}).

\bibitem[{\citenamefont{Simovic et~al.}(2003)\citenamefont{Simovic, Hucker,
  Hammel, Buchner, Ammerahl, and Revcolevschi}}]{Simovic2003}
\bibinfo{author}{\bibfnamefont{B.}~\bibnamefont{Simovic}},
  \bibinfo{author}{\bibfnamefont{M.}~\bibnamefont{Hucker}},
  \bibinfo{author}{\bibfnamefont{P.~C.} \bibnamefont{Hammel}},
  \bibinfo{author}{\bibfnamefont{B.}~\bibnamefont{Buchner}},
  \bibinfo{author}{\bibfnamefont{U.}~\bibnamefont{Ammerahl}}, \bibnamefont{and}
  \bibinfo{author}{\bibfnamefont{A.}~\bibnamefont{Revcolevschi}},
  \bibinfo{journal}{Phys. Rev. B} \textbf{\bibinfo{volume}{67}},
  \bibinfo{pages}{224508} (\bibinfo{year}{2003}).

\bibitem[{Wil()}]{Wilkins2010}
\bibinfo{note}{S. Wikins et al., unpublished results}.

\bibitem[{\citenamefont{Yamada et~al.}(1998)\citenamefont{Yamada, Lee,
  Kurahashi, Wada, Wakimoto, Ueki, Kimura, Endoh, Hosoya, Shirane
  et~al.}}]{Yamada1998}
\bibinfo{author}{\bibfnamefont{K.}~\bibnamefont{Yamada}},
  \bibinfo{author}{\bibfnamefont{C.~H.} \bibnamefont{Lee}},
  \bibinfo{author}{\bibfnamefont{K.}~\bibnamefont{Kurahashi}},
  \bibinfo{author}{\bibfnamefont{J.}~\bibnamefont{Wada}},
  \bibinfo{author}{\bibfnamefont{S.}~\bibnamefont{Wakimoto}},
  \bibinfo{author}{\bibfnamefont{S.}~\bibnamefont{Ueki}},
  \bibinfo{author}{\bibfnamefont{H.}~\bibnamefont{Kimura}},
  \bibinfo{author}{\bibfnamefont{Y.}~\bibnamefont{Endoh}},
  \bibinfo{author}{\bibfnamefont{S.}~\bibnamefont{Hosoya}},
  \bibinfo{author}{\bibfnamefont{G.}~\bibnamefont{Shirane}},
  \bibnamefont{et~al.}, \bibinfo{journal}{Phys. Rev. B}
  \textbf{\bibinfo{volume}{57}}, \bibinfo{pages}{6165} (\bibinfo{year}{1998}).

\end{thebibliography}

\end{document}